# Second Order Operators in the NASA Astrophysics Data System

Michael J. Kurtz, Roman Chyla, and the ADS Team
Center for Astrophysics | Harvard & Smithsonian
kurtz@cfa.harvard.edu, rchyla@cfa.harvard.edu, adshelp@cfa.harvard.edu

## Abstract

Second Order Operators (SOOs) are database functions which form secondary queries based on attributes of the objects returned in an initial query; they can provide powerful methods to investigate complex, multipartite information graphs.  The NASA Astrophysics Data System (ADS) has implemented four SOOs, *reviews*, *useful*, *trending*, and *similar* which use the citations, references, downloads, and abstract text.

This tutorial describes these operators in detail, both alone and in conjunction with other functions.  It is intended for scientists and others who wish to make fuller use of the ADS database.  Basic knowledge of the ADS is assumed.

## Introduction

The Second Order Operators (SOO) are a unique feature of the ADS; they have been part of the system since the very beginning of the project.  Essentially they are two-step search operations. In the first phase, the search engine retrieves a set of papers and in the second phase it will analyze those (first stage) papers and discover/collect their neighbours. The four currently implemented SOOs can analyze abstract text, references, citations, and readership information.  An example outside of ADS is Amazon's "people who bought this book also bought" feature.

Like much of the ADS, the SOOs were inspired by the work of Peter G. Ossorio whose closure operators can be viewed as iteratively performing SOOs in order to find the entire influence structure surrounding an event.  In the ADS implementation we use attributes of papers relevant for a bibliographic system, and by not iterating we measure only the most local signals in the influence network.

All current ADS SOOs take a list of articles as input, normally as returned by a query, and return a list of articles as output.  The four SOOs currently in the ADS are:

1. **Similar**.  The *similar* operator takes the text of the abstracts of the papers in the 1st order list, combines them into a single 'document', then ranks all the abstracts in the ADS by their text based similarity to this combined document, and returns the ranked list.
2. **Useful**,  The *useful* operator takes the reference lists from the papers in the 1st order list, combines them and returns this list, sorted by how frequently a referenced paper appears in the combined list.
3. **Reviews**.  The *reviews* operator takes the lists of articles which cited the papers in the 1st order list, combines them, and returns this list sorted by how frequently a citing paper appears in the combined list.
4. **Trending**.  The *trending* operator takes the lists of readers who read the papers in the 1st order list, finds the lists of papers which each of them read, combines these lists, and returns the combined list, sorted by frequency of appearance.

Extensive discussions of these operators follows.

The first of these operators to be implemented was the text based one (now called *similar*), which was included in the [original release of the ADS system](#)..  The reference and citation operators (*useful* and *reviews*) were implemented once we had the [citation data](#); the [original implementation was described in 2000](#).  We delayed releasing the usage based operator (*trending*) for a few years due to privacy concerns; [their implementation in ADS Classic was described in 2002](#).

The actual implementation of these operators in ADS Classic was not especially intuitive, and was not much used.  The main use of the SOOs in the Classic era was in the [myADS notification service](#), and the [myADS-arXiv service](#).

A detailed discussion of the ADS' [basic philosophy for search](#) shows how the SOOs functioned in one of the early experimental systems which led to the current ADS.  A brief description is on the [SOO ADS Help page](#), which describes figure 1.   Here we additionally include brief descriptions of the [topn](#), [references and citations](#) operators, and the [author and paper network](#) visualizations to clarify the presentation.

The present article is the first one to  describe in detail the use and implementation of the SOOs in the modern ADS system.

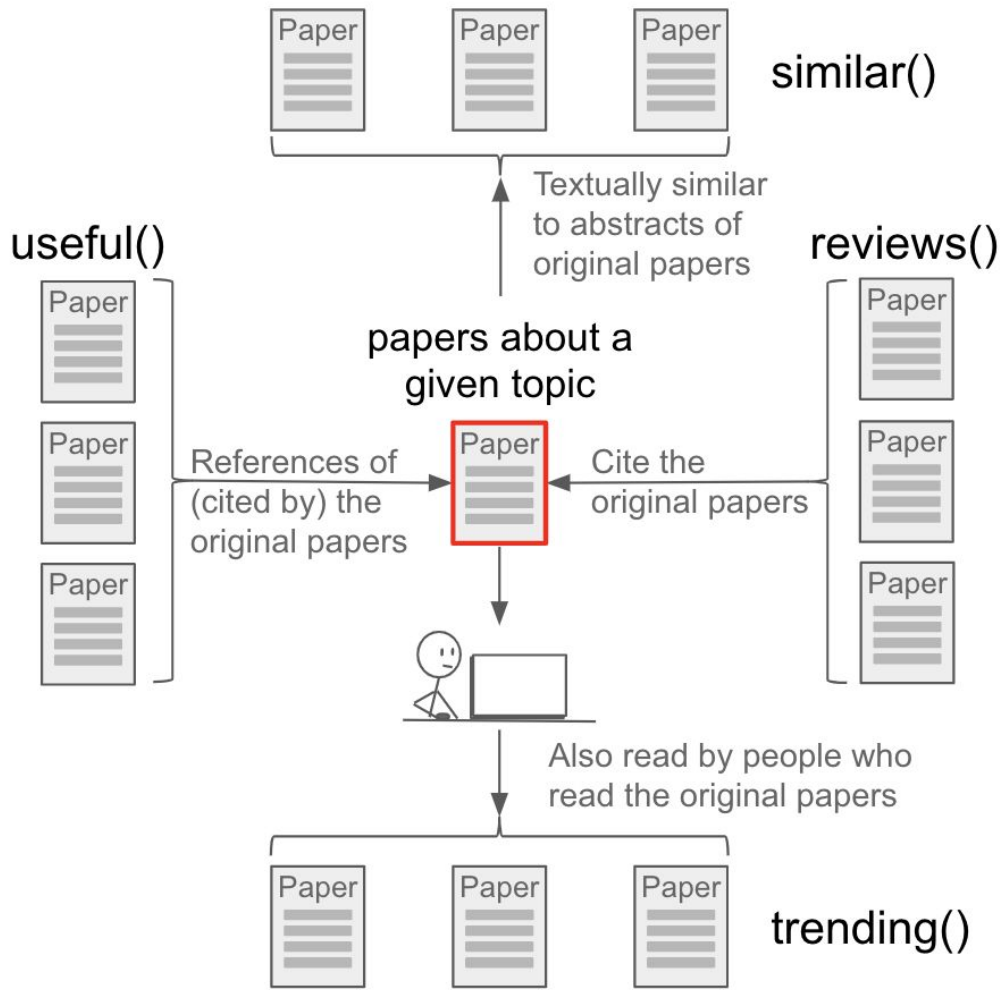

Figure 1. The four SOOs

## The *useful* operator

The *useful* operator returns the collated reference lists from a set of papers, sorted so that the paper which appears most frequently (the paper most cited by the papers in the original list) is on top. Using specific topical queries this operator changes citation ranking from being of casual, general interest to providing the researchers or students with actionable intelligence to further their scientific inquiries.

While it is certainly interesting that Lowry, et al (1951) is the most cited paper in science, this is of little practical use to an astronomer's current research. That Perdew, et al(1996) is the most cited paper in ADS, or that Riess, et al(1998) and Perlmutter, et al(1999) are still running neck and neck, with both having just passed Schlegel, et al(1998) as the most cited astronomy paper may be of even more interest to typical astronomers, but such interest is really just casual.

Garfield invented citation indexing to facilitate tracking and discovering ideas. The *useful* operator is designed to allow one to discover ideas (research articles) at the level of specificity required for high level research.

As an example we take an increasingly sophisticated set of queries. We begin by assuming a person, PostDocX, who already has a good familiarity with the topic of "exoplanet atmospheres," and keeps up to date, perhaps by using myADS. PostDocX may want to search the recent literature, to check for what other researchers in the field are currently finding useful: useful("exoplanet atmospheres" year:2019-2020) will show what papers are being cited by these articles, in citation order. Let's take it up a notch, and say that PostDocX is very familiar with the papers of the most prolific author in this field, J. Fortney, and does not need to see papers which are cited by one of Professor Fortney's papers: useful("exoplanet atmospheres" year:2019-2020) -references(author:"fortney,j") does this.

Either the full or the edited list would likely be of great interest to PostDocX, being the papers currently used in the specified sub-field, ordered by amount used (usefulness). We can take this a step further. Assume that PostDocX is really interested in atmospheric ammonia, we can modify the query: useful("exoplanet atmospheres" year:2019-2020) -references(author:"fortney,j") full:NH3 (ADS knows that ammonia is NH3).

Again, an interesting list, if one is PostDocX. Looking at a couple on the top, Wilzewski et al(2016) is a very relevant paper from the HITRAN group; while Professor Fortney has often cited HITRAN papers (references(author:"fortney,j") HITRAN), he has never cited this particular one. The publication Kasting(1982) seems to be a very interesting historical publication; we can see which papers/authors cited it: ("exoplanet atmosphere" year:2019-2020) citations(bibcode:1982JGR....87.3091K). Professor Linsky is a well established astronomer, likely known to PostDocX. Paul B. Rimmer is a 2012 PhD, now a postdoc at Cambridge working on similar problems as PostDocX (were that person not imaginary).

Were PostDocX to want to check for NH3 in the Fortney cited papers, useful("exoplanet atmospheres" year:2019-2020) references(author:"fortney,j") full:NH3 would do the trick.

We pause here to examine the role that the score parameter has played in the above. Score is a pseudo-relevance which is computed differently depending on the exact nature of the query. For the pure *useful* operator the score is just the number of citations, as described above. Once one requires "ammonia" the score changes to combine the citation information with the frequency and location of the desired word(s). Thus the papers on top have the word "ammonia" (or NH3) in the title.

*useful*, as all the SOOs, is limited to 200 papers from the inner query. If the inner query returns more than 200 papers, the papers are sorted, and the top 200 papers are used; the default sort is score for *useful*, *similar*, and *trending*, the default sort for *reviews* is citation count. When one wants to use a different sort order, and/or fewer than 200 papers for the inner query the *topn* function provides the functionality. *topn* has the format topn(N,query,sort-order). For example [topn(150,author:"sandage,a",citation_count desc)](#).

If you are familiar with the ADS user interface, you might know that you can select individual papers and then use the Explore pull-down menu *useful* command to accomplish the same thing. And the list can be then saved into an [ADS library](#) for further analysis.

Of course the libraries can be used as input also, for example using a [library of selected ADS papers](#) retrieves: [useful(docs(library/L1iIrVsLTtiA8jp984C6eA))](#). Note that the library contains 250 papers, but only the top 200 (as sorted by score) are used.

Another application of *useful* is to implement the classic bibliometric co-citation measure. If paper A cites papers B and C, then papers B and C are related via co-citation. The more often papers B and C are found in the same reference list, the stronger the co-citation bond between them. For example taking the Hypervelocity star discovery paper [Brown, et al(2005)](#), we get all papers where it is in the reference list by [citations(bibcode:2005ApJ...622L..33B)](#) and then [useful(citations(bibcode:2005ApJ...622L..33B))](#) collates that list. Brown et al. is, by construction, on top, and then the papers are listed in order of co-citation strength with that paper.

A few more short examples of *useful*.

ProfX is in charge of the Harvard Astronomy Colloquium, and is looking for speakers of local interest. ProfX can assume that colleagues suggest their own collaborators, but wishes to find additional scientists beyond this group. The query: [useful(inst:"^harvard u" year:2020 collection:astronomy property:refereed) -inst:"harvard u" year:2019-2020](#) shows, in the author facet, the people who were authors on the largest number of papers published 2019-2020 which were cited by refereed papers with Harvard first authors in 2020, but which did not have a Harvard affiliated co-author. ProfX might find the author facet contains a good list for possible speakers.

StudentX has just completed a senior thesis on ["cataclysmic variables"](#) and is looking at grad schools where continuing with this subject would be a possibility. StudentX has read articles by John Thorstensen, and wonders what he might recommend. The query: [useful(author:"^thorstensen,j.r." year:2010-2020) year:2005-2020](#) shows the recent papers and authors Professor Thorstensen has cited (as first author) the most in the last decade. Examining the lists of papers and authors one easily comes up with a list of candidates, including Kyoto, Washington, Warwick, and Ohio State (in addition to Dartmouth).

Finally let's look at an historical example. The globular clusters in the Andromeda Nebula have long been studied (Hubble(1932)); here we will look at the 100 most cited papers by papers in this field, and instead of just showing the list, we will show them graphically, using the Paper Network. The clustering algorithm separates the earlier work from the HST results, and none of these very well cited papers are newer than ten years old.

Examining this result in detail: object:m31 "globular cluster" queries both NED and SIMBAD for M31, these results are combined with ADS searching the title, abstract and keyword fields for the phrase "globular cluster". That result is sent to the *useful* operator, which sorts it by score, takes the top 200, extracts the reference lists for them, collates these lists and sorts the resulting list so the most cited papers are on top. This list is truncated to the top 100 most cited papers by *topn*, and these 100 papers are clustered using the Louvain algorithm on the network formed by the joint appearance of papers in their reference lists. The clusters are named by their most important statistically differentiating words. This link:Paper Network does all this in a couple of seconds.

## The *similar* operator

The *similar* operator is a modern implementation of the "partial match" search capability in the original ADS release;it differs from the other three SOOs (*useful*, *reviews*, *trending*) in that it does not rely on lists of connections (references, citations, readers), but uses the (abstract text of the) documents themselves. This means that, in addition to documents in the areas ADS curates (astronomy and physics), the *similar* operator will also work with the documents which the ADS simply collects, such as the math and computer science sections of arXiv.

*similar* uses the SOLR/lucene implementation of the BM25 algorithm to compare text. The collated text of all the abstracts returned by the inner query (if this is more than 200, only the top 200 are taken, according to score) are compared with each abstract in the ADS database. These documents are sorted by their degree of match, **the documents in the inner query are removed**, and the sorted list is returned, most similar paper on top.

The simplest *similar* query is of a specific paper. As an example take a recent abstract from the NED team. The "similar" link on the left of the abstract page runs the *similar* query: similar(bibcode:2020AAS...23545505M). Looking at the returned list we see that the query matched both the astronomical objects being discussed (in this case LIRGs), and the means by which they were studied (the surveys, and NED itself). The mixing of means and ends is quite typical for text based queries.

Another example: ScienceWriterX is interested in writing about recent work in optical spectrographs and knows about the work of D. Fabricant. The query author:"Fabricant,D" spectrograph finds his relevant papers. Next ScienceWriterX does the SOO: similar(author:"Fabricant,D" spectrograph) , but ScienceWriterX discovers several non-spectroscopic MMT instrument papers, as well as some older papers. Making the necessary changes and truncating to the top of the list: topn(200,similar(author:"Fabricant,D" spectrograph) -MMT year:2000-2020) shows many of the main projects on top, and also shows the importance of the SPIE Conference Series to this field. Invoking the Author Network from the Explore menu gives ScienceWriterX a graphical display of the main groups working in this field (in addition to the MMT/Fabricant group).

Now comes HistorianX, who is interested in finding the most influential work on stars in the last half century, or so. Starting with one of the most influential articles ever written, Johnson & Morgan(1953) (the fundamental paper for both MK spectral classification and UBV photometry), HistorianX selects a set of 200 similar papers, topn(200,similar(1953ApJ...117..313J)) which nearly uniformly samples the last 75 years of astronomical publications. useful(topn(200,similar(1953ApJ...117..313J))) returns the most cited papers by this group. Johnson & Morgan is (of course) on top, but browsing the list shows many of the most important research directions in the classical study of stars. Narrow band/Stromgren photometry, the Hipparchos mission, main sequence fitting of open clusters to determine/calibrate distance and age, model atmospheres, infra-red photometry all show up in the top 10.

*similar* is particularly effective in subject matter queries, but with caveats. For example the query similar("weak lensing") excludes the results of the query topn(200,"weak lensing",score desc) as those articles are used as input to the *similar* operator itself. This may, or may not be a problem, but one must be aware of it. One way to avoid this issue is to force the query to be disjoint from the result; similar("weak lensing" year:2019) year:2020 for example.

Keeping up with the literature is perhaps the main use of *similar*. Because this finds very recent publications similar to those listed in the SETI bibgroup similar(bibgroup:SETI) entdate:[NOW-7DAYS TO *], it might be used by the maintainers of the SETI bibliographic group.

A custom ranked list of recent arXiv postings can be made, similar("weak lensing" -entdate:[NOW-7DAYS TO *]) entdate:[NOW-7DAYS TO *] bibstem:"arXiv" where the -entdate command in the inner query forces the *similar* result to be disjoint, and the bibstem argument forces the result to be only arXiv postings.

The sort of query can be made for any subject matter covered by arXiv. Some examples: abell cluster redshift, BERT ELMO, "open cluster", graphene monolayer, bottom quark interaction, homology algebra, exoplanet atmospheres, author:"^accomazzi,a", citations(1998PASP..110..934K ), citations(author:"ginsparg,p" chirality), orcid:0000-0002-8035-4778

In many cases, and especially when the subject query is narrow and the time interval is short, there will be no papers which exactly match the query.  While first-order queries are useful to select papers that have an exact match to the input text, *similar* provides a way to discover related content.  *similar* finds papers which are lexically similar. This can result in serendipitous discovery.

*similar* can also be used with arbitrary text as input, in the format: *similar*("input text string",input), here for example using the first paragraph of this paper as input.  The ability to make partial match queries of input text has been in ADS from the beginning; this query duplicates one made at the announcement of the service, at the ADASS II conference in Boston in November 1992, the results can be compared with the results in figure 2 of the announcement paper.

## The *trending* Operator

*trending*, or "people who read this (these) article(s) also read," essentially provides a way to get recommendations from a custom, user selected group of people.  It differs from a most popular query (obtained simply by sorting a normal (1st order) result by read count) by only looking at what is currently popular among a chosen subset of individuals.

Before any user interacts with the readership data, ADS pre-filters the list of (anonymous) reader-article pairs to include only those readers whose readership patterns make it likely that they are active researchers, thus ensuring a high level of expertise is encoded in the returned measures.  Also ADS truncates the list to the most recent 90 days, thus ensuring currency (hence the name *trending*).

As with *useful* and *similar* the default for a broad inner query is the top 200 papers sorted by score, thus trending("weak lensing") gives exactly the same result as trending(topn(200,"weak lensing" ,score desc)).  In this case all the readers of the 200 papers from the query topn(200,"weak lensing", score desc)  are found, and all the papers each one has read in the last 90 days are collated into a list.  The list is sorted by the number or readers and returned; the most read paper (by the people who read the papers in the list of 200) is on top.

One does not need to accept the default sorting for the inner query.  For example one could use the query topn(200,"weak lensing",read_count desc) to obtain trending(topn(200,"weak lensing",read_count desc)) which gives substantially different results.  Note that 200 is the maximum number of papers allowed in the *trending* operator, so if more are input the result will be the top 200, sorted by score.

While these queries give interesting results, the full power of *trending* comes from the ability to build an idealized model person to give advice.  For example GradStudentX might want reading suggestions from an individual who has a good current knowledge of methodologies involved in weak lensing analysis. This "person" could be modeled by the sum of all the researchers who recently read the papers "weak lensing" full:b-mode year:2020, so the suggestions would be trending("weak lensing" full:b-mode year:2020).

Once GradStudentX's compound person has been created, more specific "opinions" can be requested. Essentially any ADS query can be combined with a trending query to get very specific recommended articles. Three examples which might be of interest to GradStudentX are: trending("weak lensing" full:b-mode year:2020) HST, trending("weak lensing" full:b-mode year:2020) galaxy cluster, trending("weak lensing" full:b-mode year:2020) author:"donahue,m".

It is also quite possible to model a specific individual. We begin by looking at recent 1st author articles by Igor Chilingarian author:"^chilingarian,i" year:2019-2020. We can see what people who are interested in Dr. Chilingarian's work are interested in: trending(author:"^chilingarian,i" year:2019-2020), but this does not model the author, rather those interested in the author's work.

If we instead begin with the author's reference lists references(author:"^chilingarian,i" year:2019-2020), we can build our "hive mind" model of Dr Chilingarian by trending(useful(author:"^chilingarian,i" year:2019-2020)), here we use the *useful* operator, instead of the *references* command to emphasize the fact that were the list of references larger than 200 *useful* would rank the documents by the frequency that Dr Chilingrian cited them, but references would result in the truncation to 200 documents based on the score, not the opinion of the person whose thoughts we are trying to model.

As above one can query the opinion of the model, for example about UDG (Ultra Diffuse Galaxies) trending(references(author:"^chilingarian,i" year:2019-2020)) UDG. One can also make a model of oneself, then subtract one's own work from the result to find things which might have been missed, or specifically for UDG papers. This is a useful technique, when maintaining a reference list for a paper in progress in an ADS library, in addition to the citation helper command.

There can be very substantial differences of opinion between different models on what are the most popular current articles. As an example the readers of the papers returned by the query topn(200,nucleosynthesis,score desc) form the model for the simple query trending(nucleosynthesis); the top 200 papers in this list can be visualized. Different models can be used. Beginning with the seminal B2FH paper one could make a model of persons who read recent papers on nucleosynthesis which cited this paper topn(200, citations(1957RvMP...29..547B) nucleosynthesis,date desc), giving the result trending(topn(200, citations(1957RvMP...29..547B) nucleosynthesis,date desc)). This can also be visualized. The two queries give very different results, both interesting. Which is more useful depends on the needs of the person making the query.

*trending* can be used as a current awareness search, or notification, but with substantially different time and subject properties to using *similar*. Some examples: ["exoplanet atmospheres"](), ["Machine learning"](), ["Poincare conjecture"](), ["spectral classification"](), ["population III stars"](), ["Higgs boson"](), ["Hall effect"](), etc, and all the queries for *similar*. As with *similar* one can restrict the results to recent publication dates. For example the [Pop III star query restricted to the last 90 days]().

## The *reviews* Operator

*reviews* is another operator which operates on lists of papers. Given a list of papers *reviews* returns all papers which cite any of the initial papers, with results sorted by the number of papers in the list which they cite. If the input list is a coherent set of papers on a single subject, then *reviews* tends to return review articles on top, hence the name.

The most typical use of *reviews* is to request review articles on a specific subject: [reviews("weak lensing")](). Like the other SOOs *reviews* limits the number of articles analyzed to 200, but unlike the other SOOs the default inner sort for *reviews* is citation count, not score. Thus reviews(subjectX) returns the papers which cite the largest number of the most cited papers on subjectX.

By selecting the subject of the query one can find extensive discussions of specific topics. Examples are: [exoplanet atmospheres]() vs [exoplanet -atmospheres]() or [dark matter]() vs [dark matter muon]() vs [dark matter hadron]() vs [dark matter higgs](). One can also insist that the papers be recent as with [(dark matter higgs) year:2020]().

While sorting by citation count works best for most general *reviews* queries, there are cases where sorting by score yields equivalent or better results. One example in particular is if one is interested in only newer papers, where papers have not had time to build up citations. Score includes download counts, which favors newer articles. This increases the immediacy, but lowers the signal to noise of a citation-based operator like *reviews*. Compare, for example [reviews("weak lensing" year:2015-2020)]() to [reviews(topn(200, "weak lensing" year:2015-2020, score desc))]().

Beyond its use in finding review articles *reviews* has many other uses. As an example AdministratorX wants to find the main research directions which were heavily influenced by the Panchromatic Hubble Andromeda Treasury (PHAT) project. Since PHAT has a number of meanings the query is restricted by requiring the PI be an author [reviews(full:PHAT author:dalcanton)](). This has the problem that many of the papers are by the PHAT team itself, so AdministratorX removes papers where the PI is an author [reviews(full:PHAT author:"dalcanton") -author:dalcanton](). Inspection shows this list contains many papers where the PHAT is only incidental to the result; AdministratorX then requires that the PHAT be explicitly referred to in the text  [reviews(full:PHAT author:"dalcanton") -author:dalcanton full:PHAT](#).  These papers, which are not by the PHAT team, and which make substantial use of that resource cover a range of different topics, as can be seen in the [paper network](#).

The *review*s operator can also be used to implement the classic [bibliometric coupling](#) measure.  If paper A cites paper C and paper B cites paper C then papers A and B are bibliometrically coupled, by virtue of citing the same paper, C.  The strength of the coupling is the number of papers in common in the reference lists.  For example [reviews(references(bibcode:2015ApJ...807L...2K))](#) lists the papers bibliometrically coupled with [Koda, et al(2015)](#) listed in order of coupling strength.

Bibliometric coupling provides the link strength in the Paper Network visualization.

## Synopsis: one article queries

In this section we demonstrate the use of the SOOs by looking at the results of all the operators, and simple combinations of the operators, on a single paper.  This is intended to give the reader an intuition as to how the SOOs function.  This might be improved by using a paper of the reader's own choosing; we encourage that substitution.

The ADS acts as an electro-mechanical adjunct of the memories of our human-scientist users.  Often the user knows exactly which memory is desired, such as [author:"^seager,s" mass-radius solid exoplanets](#) which returns exactly the paper [Seager, et al(2007)](#).  From the main page for this article one can read the abstract, see the figures, see its citation and download histories, learn about the (astronomical) objects discussed via SIMBAD, and get the full article from the ApJ or the preprint from arXiv.

Also from this page one may access the four lists of articles associated with this paper, References, Citations, Co-Reads, and Similar Papers. These correspond to the four SOOs *useful*, *reviews*, *trending*, and *similar*.  Each of these lists shows a different aspect of the original [Seager et al paper](#) and can be used to explore and discover connected research.

The intellectual (or idea or concept) space surrounding even a single research paper is extremely complex.  Providing tools for the scientist-user of the ADS to explore this space in a directed manner is the goal of the SOOs; here we first look at the similarities and differences between the four lists, using the graphics from the paper network visualizations.  Figure 2 shows the four figures, clicking on them will run the actual query; we truncate the lists to the top 100 papers.  (click on the X in the upper right to get the list result).

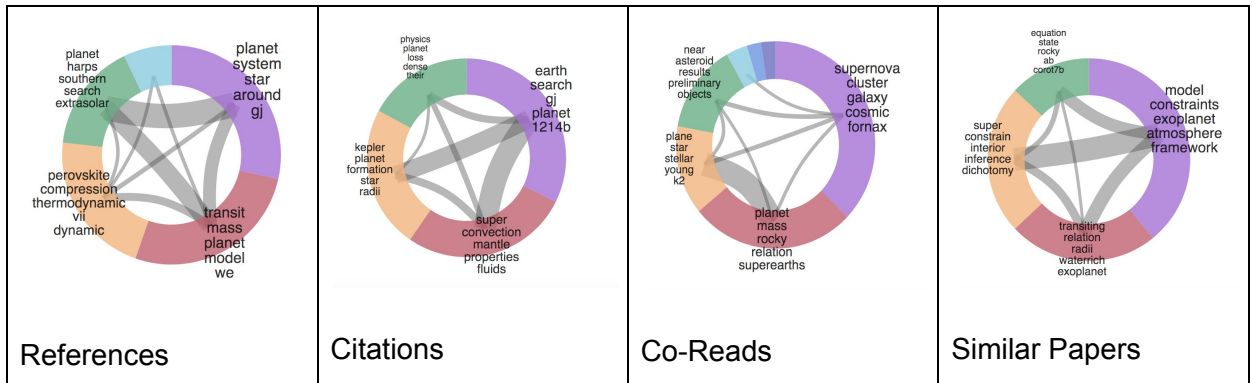

Figure 2. The four lists associated with 2007ApJ...669.1279S

The most obvious difference in the four lists is the date distribution. References all predate the 2007 publication date of the Seager, et al paper, Citations all postdate it. Co-Reads are mostly very recent, and Similar Papers can have a wide date distribution, depending on the introduction and use of words and concepts.

Just the crude subject matter clustering in the visualizations of these results makes clear that the different operators return quite different information. For a person interested in this branch of research each list can be instructive. Perhaps the most interesting factoid from these queries is that the date range of lexically similar papers only contains two papers published (a few months) before Seager, et al., both by Valencia, et al., and both are cited by 2007ApJ...669.1279S. This suggests the great timeliness and influence of these early papers.

While the graphics shown here are static, the queries behind them are dynamic, they will reflect the current state of the database, which changes daily, as papers are written, read, and cited, and as the database is corrected and enhanced. The lists and visualizations in this section (as well as all the queries in this paper) were captured on 22-June-2020; in general they will not be exactly reproduced in the future.

It is worth pointing out that two of the SOOs are paired with first order functions; *useful* with *references*, and *reviews* with *citations*. If the number of articles returned is equal to or less than 200 then the returned lists are the same, but with different sort orders (*references* alphabetical by author name, *citations* by publication date). Beyond 200 articles the SOOs truncate the input list as described above, but the first order functions perform their operations on the full list. So, for example citations(bibstem:ApJ) returns every paper known to ADS which has ever cited an *Astrophysical Journal* paper and references(bibstem:ApJ) returns every paper known to ADS referenced by an *Astrophysical Journal* paper.

In figure 3 we show each of the four SOOs operating on each of the four lists from figure 2, 16 different combinations. While some of these combinations may be more valuable than others in particular cases, and some may be more easily described (useful(references) shows the historical influences on the original research, for example) they all show different views of the research field defined by the original Seager et al paper.

Again we truncate the resulting lists to 100 papers, and we continue to encourage the substitution of [2007ApJ...669.1279S](#) with a paper in a field with which the reader is very familiar. The use of the topn(200,…) operator in the inner query forces the outer query operator to accept the inner query result; failure to do this can yield meaningless results, as with e.g. reviews(trending).

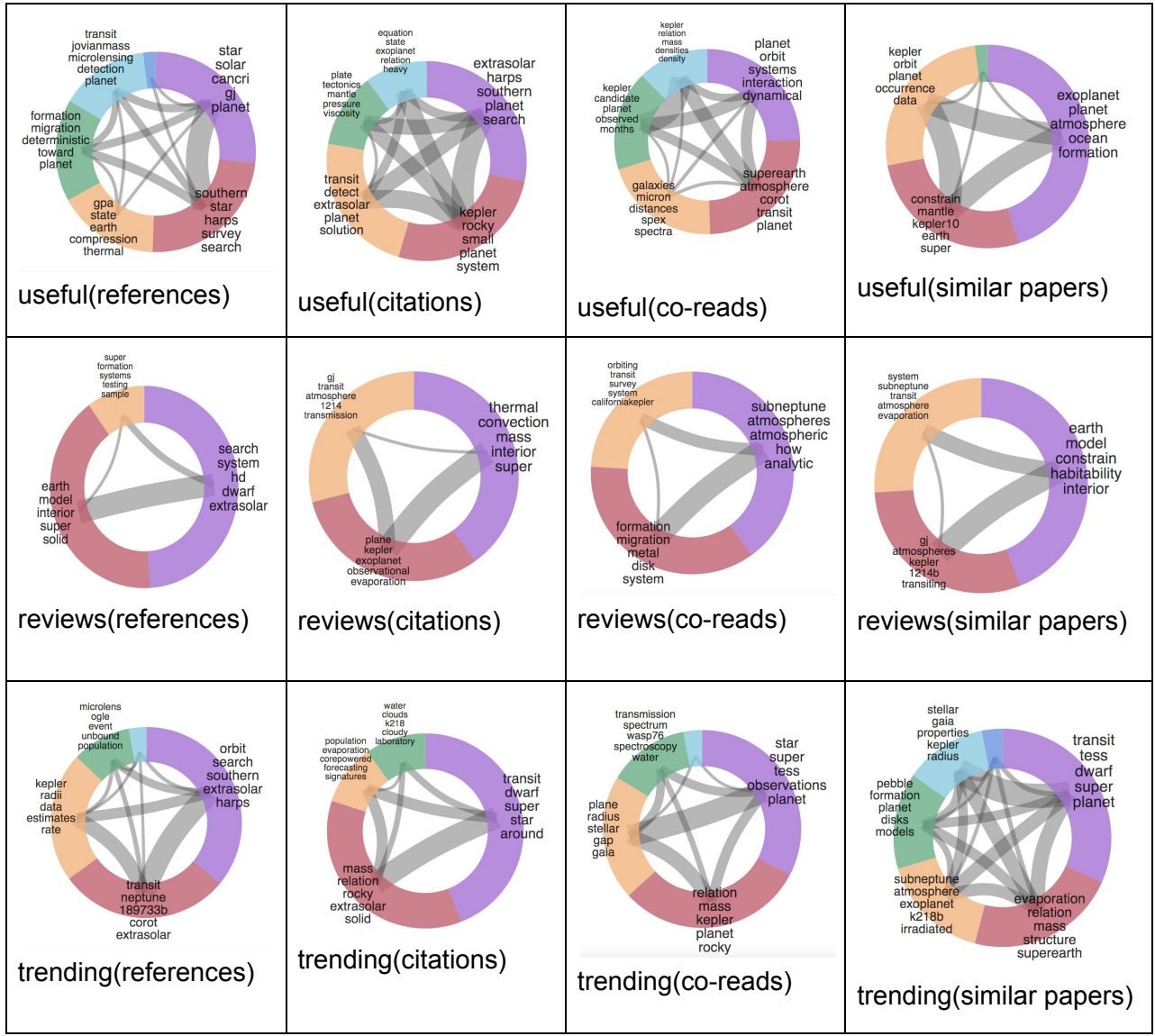

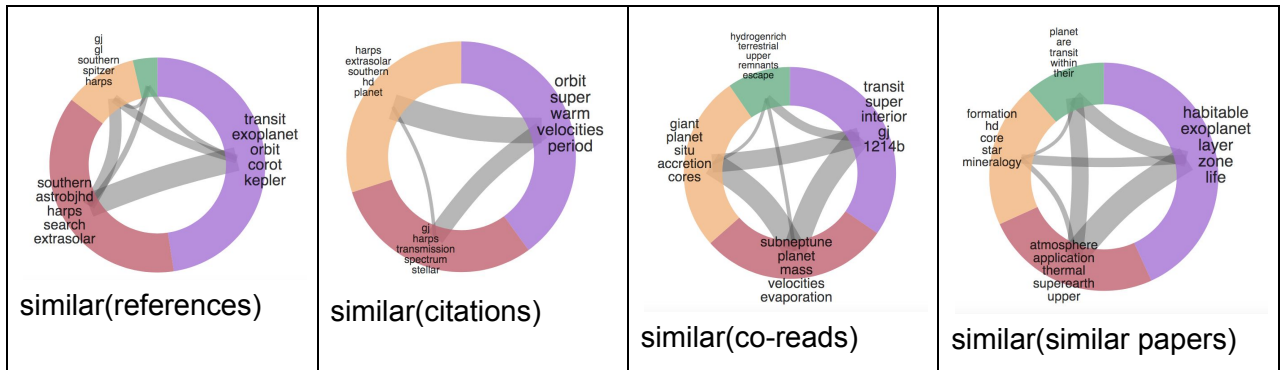

Figure 3. The four SOOs applied to each of the four lists associated with 2007ApJ...669.1279S

## Discussion

Mastering the complexities of scholarly thought is a key aspect of advanced education. Understanding how to solve partial differential equations, for example, requires several months of study, following a few years of preparation (calculus, ordinary differential equations, etc.).

Learning to take advantage of the information contained in scholarly information systems requires substantial effort. Their existence is new, but their use is already essential. Just as massive data sets of measurements, such as from the Large Hadron Collider or the Rubin Observatory require highly sophisticated, newly developed methods for their effective exploitation; the new, massive text datasets require new techniques to be invented and learned.

Compared with measurement data text datasets are small. The raw data stream from the LHC can be a petabyte per second; just the public, reduced datasets from one LHC experiment (CMS) are a few hundred PB. All the text ever published by the *Physical Review* is about 100GB, a few million times smaller than the CMS public data, which gives some indication of the concentration and complexity of textual data.

The ADS (and other similar services) can be viewed as a large multipartite network, with the links and nodes representing papers, authors, readers, keywords, references, citations, organizations, astronomical objects, …. The number of possible paths through a network scales as a factorial of the number of nodes and links; for the ADS a reasonable guesstimate is $10^{1,000,000,000}$ plus or minus a factor of two in the exponent. This dwarfs the largest astronomical measures, such as the number of particles ($10^{80}$) or planck volumes ($10^{160}$) in the universe.

These data are fundamentally new, and are transforming society, not just scholarship. The five largest industrial corporations (MSFT, APPL, AMZN, GOOG, FACB) all have large R&D efforts in the areas of machine learning, text understanding, and network analysis. While the principal goal of commercial activity is to sell widgets and tickets, etc. the technologies involved are deeply changing society, the way people interact and learn.

The role of the ADS is not to be the authoritative source of facts ([what is the value of the Hubble constant?](#)), rather it is to collaborate with researchers to enable discoveries.  The rapid increase in the velocity and volume of disseminated ideas requires intelligent management both to discover the relevant information and to prevent it from being buried in a sea of the (for a particular purpose) irrelevant.

These developments are leading to a new mode of scholarly thought, caused by the [emergent](#) behavior of groups of individual scientists responding to the near real time knowledge of the thoughts and actions of the ensemble of all their peers.  While scientific thought has always been cumulative (we *are* standing on the [shoulders of giants](#)), we are in the midst of a phase transition in how we think and store our thoughts, similar (but much faster) to the revolution caused by Johannes Gutenberg's invention of the printing press.

The SOOs are part of the set of tools which the ADS provides its users to manage and control this information flow; their use is clearly collaborative.  This article is intended to give researchers insight and intuition into how to use them to enable their research.  Working through this may seem a bit of a slog, but probably no more than a single chapter in a textbook on partial differential equations, methods of mathematical physics, or numerical analysis.